\documentclass[useAMS,usegraphicx]{mn2e}
\usepackage{epstopdf}
\usepackage{mathrsfs}

\title{A Shack-Hartmann wavefront sensor projected onto the sky with reduced focal anisoplanatism}
\author[T. Butterley et al.]{T.~Butterley,\thanks{E-mail:timothy.butterley@durham.ac.uk}
        G.D.~Love, 
        R.W.~Wilson, 
        R.M.~Myers 
        and T.J.~Morris \\
University of Durham, Dept. of Physics, Rochester Building, South Road, Durham, DH1 3LE, UK}

\begin{document}

\newcommand{\ud}{\mathrm{d}}
\newcommand{\curlyD}{\mathscr{D}}

\date{Accepted. Received.}

\pagerange{\pageref{firstpage}--\pageref{lastpage}} \pubyear{2006}

\maketitle

\label{firstpage}

\begin{abstract}
\noindent A method for producing a laser guide star wavefront sensor for adaptive optics with reduced focal anisoplanatism is presented. A theoretical analysis and numerical simulations have been carried out and the results are presented. The technique, named SPLASH (Sky-Projected Laser Array Shack-Hartmann), is shown to suffer considerably less from focal anisoplanatism than a conventional laser guide star system. The method is potentially suitable for large telescope apertures ($\sim$ 8 m), and possibly for extremely large telescopes.
\end{abstract}


\section{INTRODUCTION}
\label{sect:intro}

Laser guide stars (LGSs) \cite{foy85} solve some of the drawbacks of conventional adaptive optics (AO) systems employing natural guide stars (NGSs), most notably sky coverage limitations. However, unlike NGS AO systems, LGSs suffer from focal anisoplanatism (FA) or the ``cone effect'' \cite{parenti94}. This effect becomes more severe with increasing telescope aperture size, meaning that a single LGS on an extremely large telescope (ELT) would be unusable. FA can be at least partially overcome using multiconjugate adaptive optics (MCAO) \cite{beckers88}, in which measurements from multiple LGSs are combined. In this paper we outline an alternative to a conventional LGS in which FA is significantly reduced for a single-conjugate AO system. The technique, which we call \emph{SPLASH} \emph{(Sky Projected Laser Array Shack-Hartmann)}, is a pseudo-reverse of a conventional Shack-Hartmann wavefront sensor (WFS). Unlike a conventional LGS system, the atmosphere is sensed on the \emph{upward path} of the laser. An array of converging beams is launched from the primary mirror of the telescope to produce an array of Shack-Hartmann spots projected onto the sky, which are then imaged by the whole telescope.

A number of alternatives to ``conventional'' LGS wavefront sensing have been suggested. Baharav et al. \shortcite{baharav94,baharav96} proposed the creation of a fringe pattern in the atmosphere which is analysed by a Shack-Hartmann wavefront sensor. A disadvantage is that very high laser powers are required, although this has recently been ameliorated by the suggested adaptation of a pyramid wavefront sensor \cite{ribak01}. A proposal by Lloyd-Hart et al. \shortcite{lloydhart} involves producing a number of images of different planes in the atmosphere as the laser propagates through a focus. These images are then used in a phase diversity wavefront sensor. Angel \shortcite{angel} also proposed a system of dynamically refocusing the laser spot so as to effectively increase the integration time. Kellner et al. \shortcite{kellner04} proposed using Bessel beams as pseudo-inverse guide stars (PIGS). Ribak and Ragazzoni \shortcite{ribak04} also proposed a method of reducing laser spot elongation using distributed launch optics. All of these systems share the common characteristic with the conventional LGS system that the aberrations are sensed during the return downwards path of the laser, unlike the scheme proposed here, and they also suffer from focal anisoplanatism.

A further alternative, $P^{4}$ (Projected Pupil Plane Patterns) \cite{buscher} shares similarities with SPLASH. A laser beam is expanded to fill the pupil of the telescope and is propagated upward through the atmosphere as a parallel beam. The beam cross-section is imaged at different altitudes and the wavefront distortion is determined by comparing the intensity distributions. This method does not suffer from focal anisoplanatism and, like SPLASH, the aberrations are sensed on the upward path. However, $P^{4}$ potentially suffers from two key drawbacks: the observation layers need to be separated sufficiently in altitude for the beam intensity distribution to evolve, and the measurements may be distorted by non-uniform scattering from the atmosphere. These limitations are not applicable to SPLASH.

The basic concept of SPLASH was first presented in Love et al. \shortcite{love04}, but has much in common with Tscherning aberrometry -- a technique used in ophthalmology to measure optical aberrations in the human eye. This was first described by Tscherning \shortcite{tscherning1894} and has become more widely adopted in recent years, eg. Mrochen et al. \cite{mrochen00}. Here we present results of a theoretical analysis and a closed-loop numerical simulation to validate the SPLASH technique. In Sect.~\ref{sect:description} we outline the concept and describe the advantages and problems associated with the technique. We present our theoretical analysis and its results in Sect.~\ref{sect:theoretical}, and a description of our numerical model and its results in Sect.~\ref{sect:closed_sim}. In Sect.~\ref{sect:conc} we summarise our results and outline the remaining problems not addressed in this paper.

\section{DESCRIPTION OF SPLASH}
\label{sect:description}

This section describes the key principles of the SPLASH technique of wavefront sensing and outlines the advantages and disadvantages associated with the technique.

\subsection{SPLASH concept}

\begin{figure*}
\centering
\includegraphics[width=12cm]{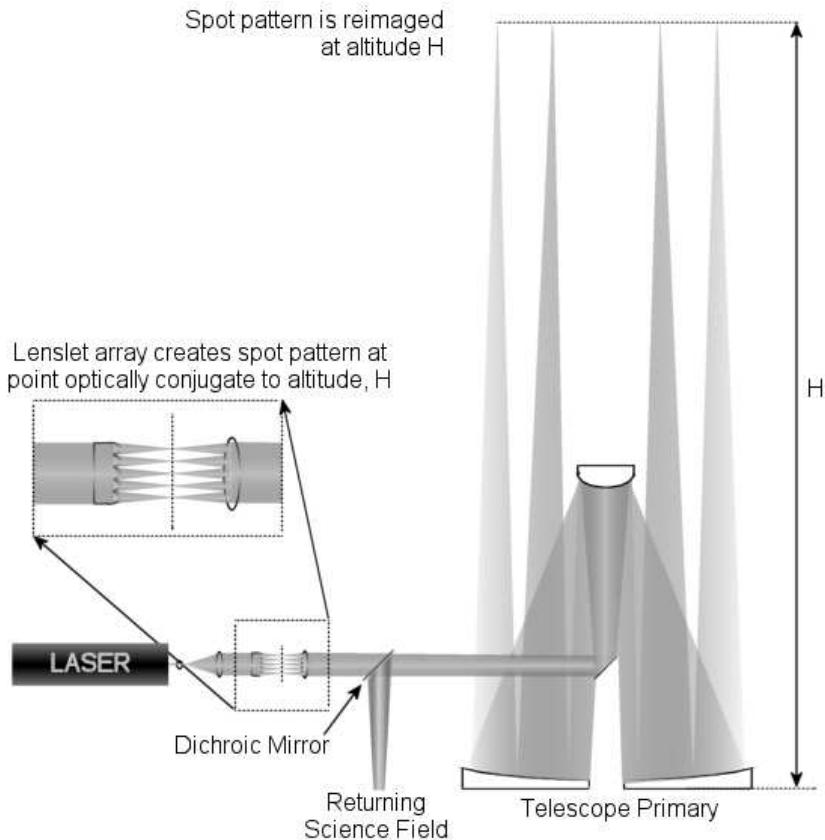}
\caption{Concept of SPLASH, showing the upward passage of the beams.
A possible optical implementation is shown whereby the laser is launched via a lenslet array. This is only a conceptual diagram and not a formal optical design - practical implementation may require a rather different approach for reasons discussed in the main text. The size of each of the converging beams is $\sim r_{0}$ although we have only shown four here for clarity. Furthermore, we have shown the beams as converging to a point, whereas in reality they would be broadened by diffraction. See text for more discussion.}
\label{fig:splash}  
\end{figure*}

SPLASH wavefront sensing is a pseudo-inverse of conventional LGS/Shack-Hartmann wavefront sensing in which the atmosphere is sensed on the upward path of the laser beam instead of the return path. The basic premise, illustrated in Fig.~\ref{fig:splash}, is to project an array of converging laser beams, each of size $\sim r_{0}$ (where $r_{0}$ is the Fried parameter), from the primary mirror of the telescope to form an array of spots on the sky. The position of each spot on the sky depends on the local (subaperture) wavefront gradient. The spots are imaged through the full telescope aperture, so the position of the final image of each spot will be altered by the global (full aperture) tilt. Hence the position of each spot image gives a measure of the local tilt minus the global tilt -- exactly the same quantity as is measured in a conventional Shack-Hartmann wavefront sensor (WFS) when used with a laser beacon. This assumes that the angular size of the spot pattern on the sky is smaller than the isokinetic patch (the angle over which the wavefront tilt is isoplanatic). 

The technique could be implemented with either a Rayleigh or a sodium beacon, although implementation of a sodium layer SPLASH system presents more of a challenge in focusing the beams than the Rayleigh version, due to the extremely long focal length. Range gating is essential, for both the Rayleigh and sodium versions of SPLASH, in order to allow the beacons to be imaged without being swamped by backscattered light from lower altitudes.

\subsection{Focal anisoplanatism}
\label{sect:SPLASH_FA}

The main proposed advantage of a SPLASH LGS is that it suffers considerably less from FA than a conventional LGS. However, it is still affected by FA and, as a result of the unusual configuration, the FA effects are manifested in a different way to that seen in a conventional LGS/Shack-Hartmann WFS combination.

\begin{figure*}%[ht]
\centering
\includegraphics[width=12.6cm]{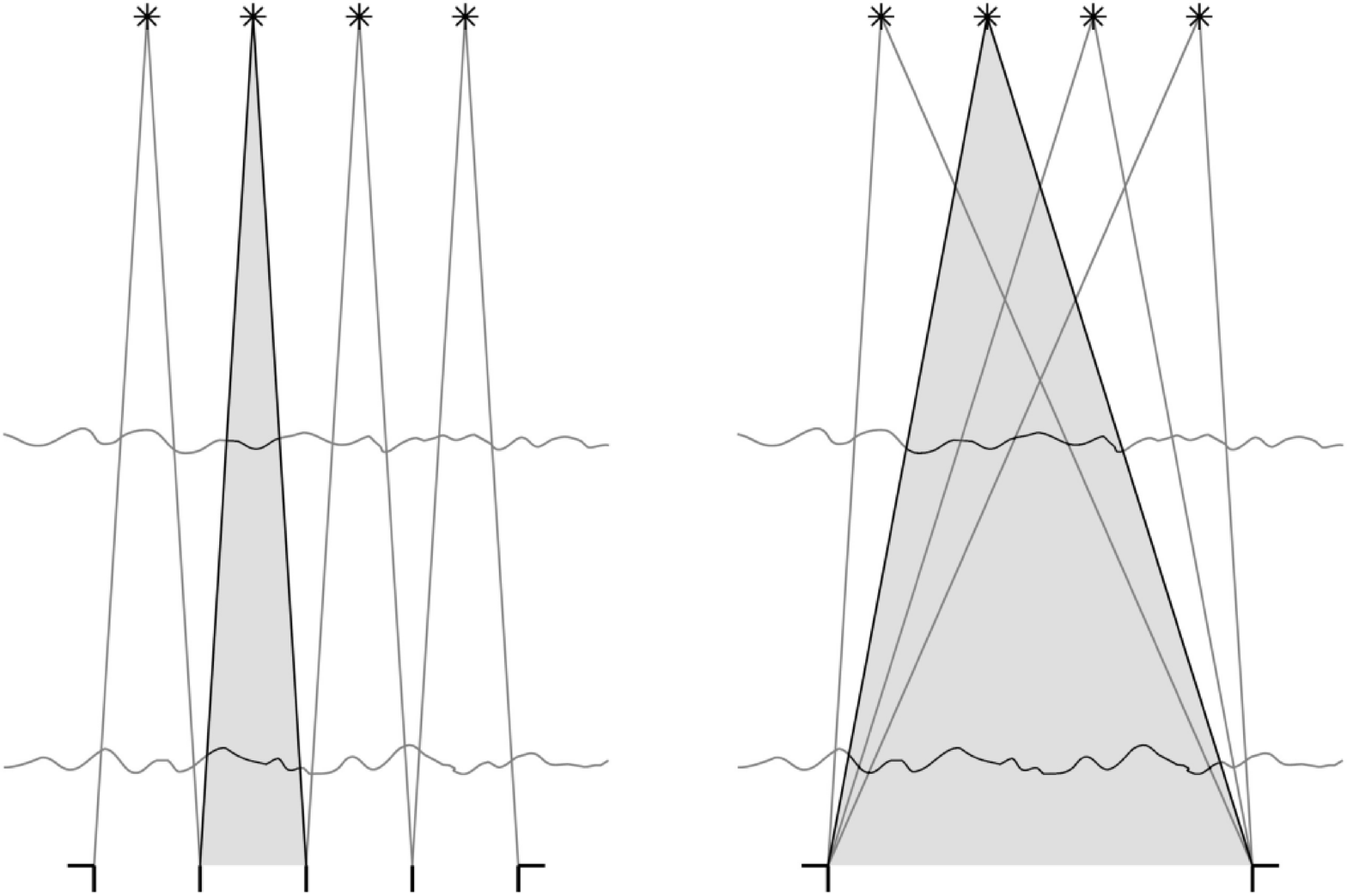}
\caption{\emph{Left:} The upward paths followed by the beams. Each beam samples the atmosphere above its own subaperture, and each beam is affected separately by focal anisoplanatism. \emph{Right:} The downward paths taken by the light from each spot. The spots are observed through the full telescope aperture so the light does not pass through the same section of atmosphere as the upward-propagating (wavefront sensing) beams. The upward-propagating beams are affected by the local wavefront gradient, and the downward beams by the global gradient (corrupted by focal and angular anisoplanatism). The paths from one subaperture are darkened to show the regions of the atmosphere encountered by light from that subaperture. The horizontal lines indicate turbulent layers.}
\label{fig:geometry}
\end{figure*}

SPLASH is affected by FA twice, on the upward and downward paths of the light. On the upward path, where the wavefront is sensed, the local gradient measured by each beam will have a slight error due to FA -- see Fig.~\ref{fig:geometry}. As each spot is then imaged from a different position in the sky, the return paths of the light from the spots each sample the atmosphere differently so that the global tilt on each subaperture centroid is separately corrupted by a combination of focal and angular anisoplanatism.

There is an additional error on the upward path due to a ``lever arm effect''. The distance a spot moves due to a local wavefront tilt depends on the altitude at which the tilt is applied -- a ground layer tilt will cause a larger spot motion than a tilt of the same magnitude applied higher up. This is not expected to be a significant problem for closed loop operation, especially for a sodium LGS system, although it will have the effect of slightly reducing the system bandwidth.

\subsection{Effect of diffraction on SPLASH}

Figures \ref{fig:splash} and \ref{fig:geometry} assume geometrical optics but in reality the SPLASH beams will be affected by diffraction, resulting in a finite spot size on the sky. The minimum size of the spots depends on the size of the launch subapertures, the laser wavelength, the laser beam intensity distribution and the beacon altitude, with the seeing resulting in a further increase in the size of the spots. Thus the spots could potentially overlap on the sky making centroiding difficult or impossible. For the sodium beacon case the longer beam propagation distance will result in larger sky spots than in the Rayleigh beacon case, worsening the problem of overlapping spots.

For beams with a Gaussian intensity profile, minimum spot size on sky would be achieved by launching the beams from subapertures of size $\sim$ 2-3 $r_0$. This would limit the wavefront sampling, meaning that SPLASH would be unsuitable for high-order wavefront sensing. However, such a system would be ideal for adaptive optics in the infrared. Launching Gaussian beams would not be possible using the approach illustrated in figure \ref{fig:splash} but could be achieved using, for example, holographic techniques.

The problem of overlapping spots could also be reduced by implementing a method of time-interleaving the subapertures such that only a subset of the beams would be launched at any given time. All of the spots would be projected and imaged in each complete cycle of the system, but the wavefront-sensing would be divided into sub-cycles with each sub-cycle involving a combination of subapertures chosen to avoid overlapping of the spots.

Aside from the issue of overlapping spots, diffraction effects may serve to partially remove the effects of FA on the upward paths of the beams. The broadened beams will sample higher-altitude turbulent layers more fully than the idealised geometrical-optics ``cones'' shown in figures \ref{fig:splash} and \ref{fig:geometry}.

\subsection{Effect of turbulence on return path}

Return path tip/tilt has already been discussed but some consideration needs to be given to higher order aberrations. The subapertures in a conventional Shack-Hartmann WFS are of similar size to the SPLASH launch subapertures, i.e. $\sim r_0$. Thus the wavefront aberration across each subaperture is dominated by low spatial frequencies and the subaperture images are generally not speckled. In SPLASH the sky spots are imaged through the full telescope aperture (with a many-$r_0$ diameter) and the turbulence-induced aberrations on the downward path can contain significant high spatial frequency aberrations. As a result the point spread functions (PSFs) could become speckled, and as the amount of speckling increases the centroid of the PSF can be expected to become poorly correlated with the local wavefront tilt. This effect is known as ``centroid anisoplanatism'' \cite{yura85,churnside85}, and becomes significant for values of $D/r_0$ greater than about 10.

The speckling effect could be reduced by masking off a large portion of the aperture so that the spot pattern would be imaged through a much smaller aperture. This is not really a viable solution, however, because it would lead to an increase in tip/tilt anisoplanatism (i.e. differences in global tip/tilt on the return paths from the sky spots).

In a closed-loop system the return-path aberrations would be removed by the wavefront corrector, so only the residual wavefront error would contribute to centroid anisoplantism. However, if the spots were too badly speckled to begin with it might be impossible to close the loop.

\subsection{Laser power requirements}

We do not present any formal laser power calculations here, however the power requirements of a SPLASH system can be expected to be comparable to those of a conventional single-LGS system. In a traditional LGS system the light from a single beacon is split between many WFS subapertures. In a SPLASH system, although a separate beacon is launched from each subaperture, each beacon is imaged through the full telescope aperture and the brightness of the beacons can be reduced accordingly. Therefore, providing the beams launched from the different subapertures all have the same intensity, the increase in collecting area balances the increase in the number of beacons. Launch methods in which the beam brightness varies between subapertures, such as that illustrated in Fig.~\ref{fig:splash}, will require higher laser power.

We now present a theoretical analysis of SPLASH, taking into account the effect of FA on the upward path but neglecting the effects of diffraction and return path turbulence.

\section{THEORETICAL ESTIMATE OF THE EFFECT OF FOCAL ANISOPLANATISM ON SPLASH}
\label{sect:theoretical}

This section describes a modal analysis of a SPLASH WFS, giving an estimate of the effect of FA on the upward (wavefront-sensing) path of the light. The effects of turbulence on the return path are excluded (but are considered later in section \ref{sect:closed_sim}).

\begin{figure*}%[ht]
\centering
\includegraphics[width=17cm]{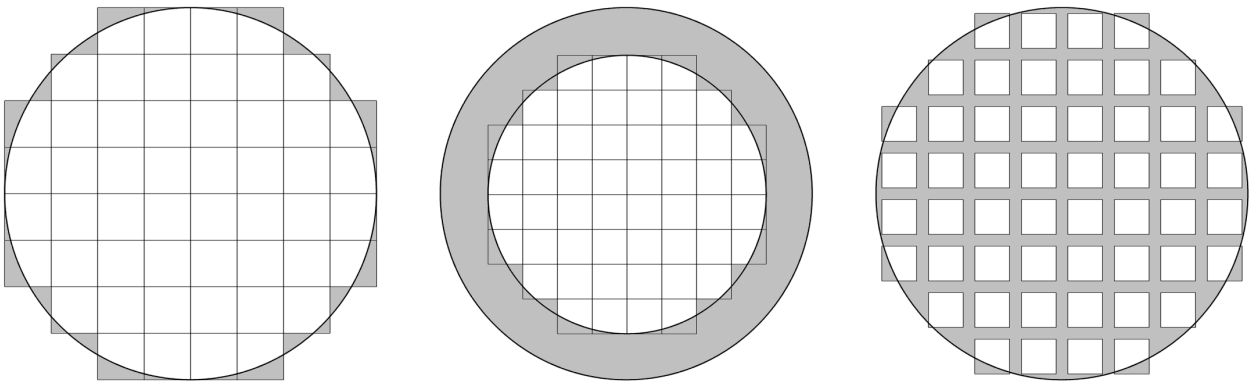}
\caption{Illustration of how the telescope pupil is projected onto a turbulent layer at height $h = H/4$. Areas shaded grey are not sensed.
	\emph{Left:} Natural guide star - light is parallel so the pupil is fully sampled at all altitudes;
	\emph{Middle:} Conventional LGS - as a result of FA, the entire pupil is projected onto a smaller circle as altitude increases;
	\emph{Right:} SPLASH - each subaperture is projected onto a smaller square with increasing altitude as a result of FA, but the spacing of the subapertures remains the same.}
\label{fig:FA_projection}
\end{figure*}

Zernike polynomials are a convenient basis set for a modal analysis of wavefront correction on a circular aperture. We will use the same conventions for normalisation and numbering of Zernike modes as described by Noll \shortcite{noll76}. The phase distortion, $\phi(R\mathbf{r})$, across a circular aperture can be expressed in terms of Zernike polynomials, $Z_j$, as
\begin{equation}
	\label{eq:zerns}
	\phi(R\mathbf{r}) = \sum^{\infty}_{j=1} a_j Z_j(\mathbf{r}) ,
\end{equation}
where $R$ is the radius of the aperture, $\mathbf{r}$ is the spatial coordinate normalised to unit radius and $a_j$ are the Zernike coefficients, given by
\begin{equation}
	\label{eq:zerncoeff}
	a_j = \int \phi(R\mathbf{r}) Z_j(\mathbf{r}) W(\mathbf{r}) \ud\mathbf{r} ,
\end{equation}
where $W(\mathbf{r})$ is the pupil function. We explicitly exclude the aperture-averaged phase (piston) from the summation in equation \ref{eq:zerns}. If the first $N$ Zernike modes could be perfectly corrected, the residual wavefront distortion would then be
\begin{equation}
	\phi(R\mathbf{r}) = \sum^{\infty}_{N+1} a_j Z_j(\mathbf{r}) .
\end{equation}
Providing the Zernike modes are normalised as described by Noll \shortcite{noll76}, the mean square residual phase error across the aperture can be written as
\begin{equation}
	\label{eq:nollvar}
	\sigma^2_\phi = \langle \phi^2 \rangle - \sum^{N}_{j=1} \langle |a_j|^2 \rangle.
\end{equation}
A wavefront sensor gives estimates $b_j$ of the first $N$ Zernike coefficients $a_j$. If these modes were corrected as accurately as they could be measured, the residual phase variance would be
\begin{eqnarray}
	\label{eq:noFAvar}
	\sigma^2_\phi & = & \sum^{N}_{j=1} \langle | a_j - b_j |^2 \rangle + \sum^{\infty}_{j=N+1} \langle a_j^2 \rangle \nonumber \\
	& = & \sum^{N}_{j=1} \langle | a_j^2 + b_j^2 - 2 a_j b_j | \rangle + \sum^{\infty}_{j=N+1} \langle a_j^2 \rangle \nonumber \\
	& = &  \sum^{\infty}_{j=1} \langle a_j^2 \rangle + \sum^{N}_{j=1} \left( \langle b_j^2 - 2 a_j b_j \rangle \nonumber \right) \\
	& = &  \sum^{\infty}_{j=1} \langle a_j^2 \rangle + \sum^{N}_{j=1} \langle b_j^2 \rangle - 2 \sum^{N}_{j=1} \langle a_j b_j \rangle .
\end{eqnarray}
The variances of the modal coefficients $\langle a_j^2 \rangle$ are given by the leading diagonal of Noll's Zernike covariance matrix. Thus to predict the performance of the WFS, we need to know the variances of the estimated modal coefficients $\langle b_j^2 \rangle$ and the covariances between the estimated modal coefficients and the actual modal coefficients $\langle a_j b_j \rangle$. These variances can be calculated for a WFS with an NGS using techniques described by Wilson \& Jenkins \shortcite{wilson96} and Cubalchini \shortcite{cubalchini79}.

The phase gradient averaged over subaperture $i$ of a Shack-Hartmann WFS is given by
\begin{equation}
	g_i = \frac{\lambda}{\pi D} \sum^{\infty}_{j=1} a_j \int_{\mbox{\scriptsize subaperture }i} \nabla Z_j (\mathbf{r}) \ud \mathbf{r}
\end{equation}
where $D$ is the telescope aperture diameter and $\lambda$ is the wavelength (the factor $\lambda / \pi D$ scales the phase tilt from units of radians of phase per telescope radius to radians of angle). There are two orthogonal phase gradients for each subaperture (commonly referred to as tip and tilt). This equation can be written in matrix form as
\begin{equation}
	\mathbf{g} = \frac{\lambda}{\pi D} \curlyD^{\infty} \mathbf{a}
\end{equation}
where $\mathbf{g}$ and $\mathbf{a}$ are the vectors of subaperture gradients and modal coefficients respectively and $\curlyD^{\infty}$ is the matrix of ($x$ and $y$) subaperture-averaged derivatives of a large number of Zernike functions. Note that, due to the Zernike normalisation used, the elements of $\curlyD^{\infty}$ are in units of radians of phase per telescope radius. The modal coefficient estimates $\mathbf{b}$ are usually found using
\begin{equation}
	\mathbf{b} = \frac{\pi D}{\lambda} \curlyD^{-1} \mathbf{g}
\end{equation}
where $\curlyD^{-1}$ is the least squares inverse of a $\curlyD$-matrix containing a small number of Zernike modes (only as many as the WFS is of sufficiently high order to measure). The modal covariance matrix for the WFS is given by
\begin{equation}
	\mathbf{C} = \left( \frac{\pi D}{\lambda} \right) ^2 \curlyD^{-1} \mathbf{G} (\curlyD^{-1})^T
\end{equation}
where $\mathbf{G}$ is a matrix of tip and tilt covariances between subapertures (i.e. $\mathbf{G}_{il}=\langle g_i g_l \rangle$). Each element of the matrix $\mathbf{C}$ is a covariance between two Zernike modes, i.e. $\mathbf{C}_{jk}= \langle b_j b_k \rangle$. For a perfect wavefront sensor the matrix $\mathbf{C}$ would be equal to Noll's matrix (i.e. $\langle b_j b_k \rangle = \langle a_j a_k \rangle$).

Up to this point the analysis described has been for a Shack-Hartmann WFS viewing an NGS (equivalent to a conventional LGS or SPLASH system with turbulence only at the ground i.e. with no FA). We are interested in the effect of FA on the performance of a SPLASH WFS, so we now introduce this into the analysis. We consider an atmosphere consisting of just one turbulent layer, although extending the analysis to include multiple layers is straightforward. We take the height of the layer above the ground to be $h$ and the beacon height to be $H$. Figure \ref{fig:FA_projection} shows how FA affects the sampling of an atmospheric layer. We define $b^{\prime}_j$ to be the SPLASH estimate of Zernike coefficient $a_j$ in the presence of FA (note that $b_j$ still represents the estimate $a_j$ in the absence of FA). Thus, with reference to Equation \ref{eq:noFAvar}, the residual phase variance is now given by
\begin{eqnarray}
	\label{eq:withFAvar}
	\sigma^2_\phi =  \sum^{\infty}_{j=1} \langle a_j^2 \rangle + \sum^{N}_{j=1} \langle b_j^{\prime 2} \rangle - 2 \sum^{N}_{j=1} \langle a_j b_j^\prime \rangle .
\end{eqnarray}

We define $g_i^\prime$ to be the phase gradient averaged over the projection of subaperture $i$ onto the turbulent layer, given by
\begin{equation}
	g_i^\prime = \frac{\lambda}{\pi D} \sum^{\infty}_{j=1} a_j \int_{\mbox{\scriptsize subaperture }i \mbox{\scriptsize \hspace{2pt}with FA}} \nabla Z_j (\mathbf{r}) \ud \mathbf{r}.
\end{equation}
Then $\mathbf{g}^\prime$ is the vector of FA-affected subaperture gradients, given by
\begin{equation}
	\mathbf{g}^\prime = \frac{\lambda}{\pi D} \curlyD^{\prime \infty} \mathbf{a}
\end{equation}
where $\curlyD^{\prime \infty}$ is the matrix of  a large number of Zernike functions averaged over FA-projected subapertures (see Figure \ref{fig:FA_projection}).  Two more modal covariance matrices can now be calculated:
\begin{eqnarray}
	\mathbf{C}^{\prime} & = & \left( \frac{\pi D}{\lambda} \right) ^2 \curlyD^{\prime -1} \mathbf{G}^{\prime} (\curlyD^{-1})^T \\
	\mathbf{C}^{\prime \prime} & = & \left( \frac{\pi D}{\lambda} \right) ^2 \curlyD^{\prime -1} \mathbf{G}^{\prime \prime} (\curlyD^{\prime -1})^T
\end{eqnarray}
where the elements of these $\mathbf{G}$- and $\mathbf{C}$-matrices are
\begin{eqnarray}
	\mathbf{G}^{\prime}_{il} & = & \langle g_i g^\prime_l \rangle \\
	\mathbf{G}^{\prime \prime}_{il} & = & \langle g^\prime_i g^\prime_l \rangle
\end{eqnarray}
\begin{eqnarray}
	\mathbf{C}^{\prime}_{jk} & = & \langle b_j b^\prime_k \rangle \\
	\mathbf{C}^{\prime \prime}_{jk} & = & \langle b^\prime_j b^\prime_k \rangle .
\end{eqnarray}

We make the assumption that the performance of the wavefront sensor is essentially perfect in the absence of FA, i.e. $a_j=b_j$. Since we are interested in investigating the effect of FA on SPLASH this is a reasonable assumption, providing that the number of reconstructed modes is limited sufficiently to avoid significant fitting error. Thus the diagonals of the matrices $\mathbf{C}^{\prime}$ and $\mathbf{C}^{\prime \prime}$ are equal to the values $\langle a_j b_j^\prime \rangle$ and $\langle b_j^{\prime 2} \rangle$ respectively, which are the unknowns required in Equation \ref{eq:withFAvar}.

\begin{figure*}
\centering
\includegraphics[width=8cm]{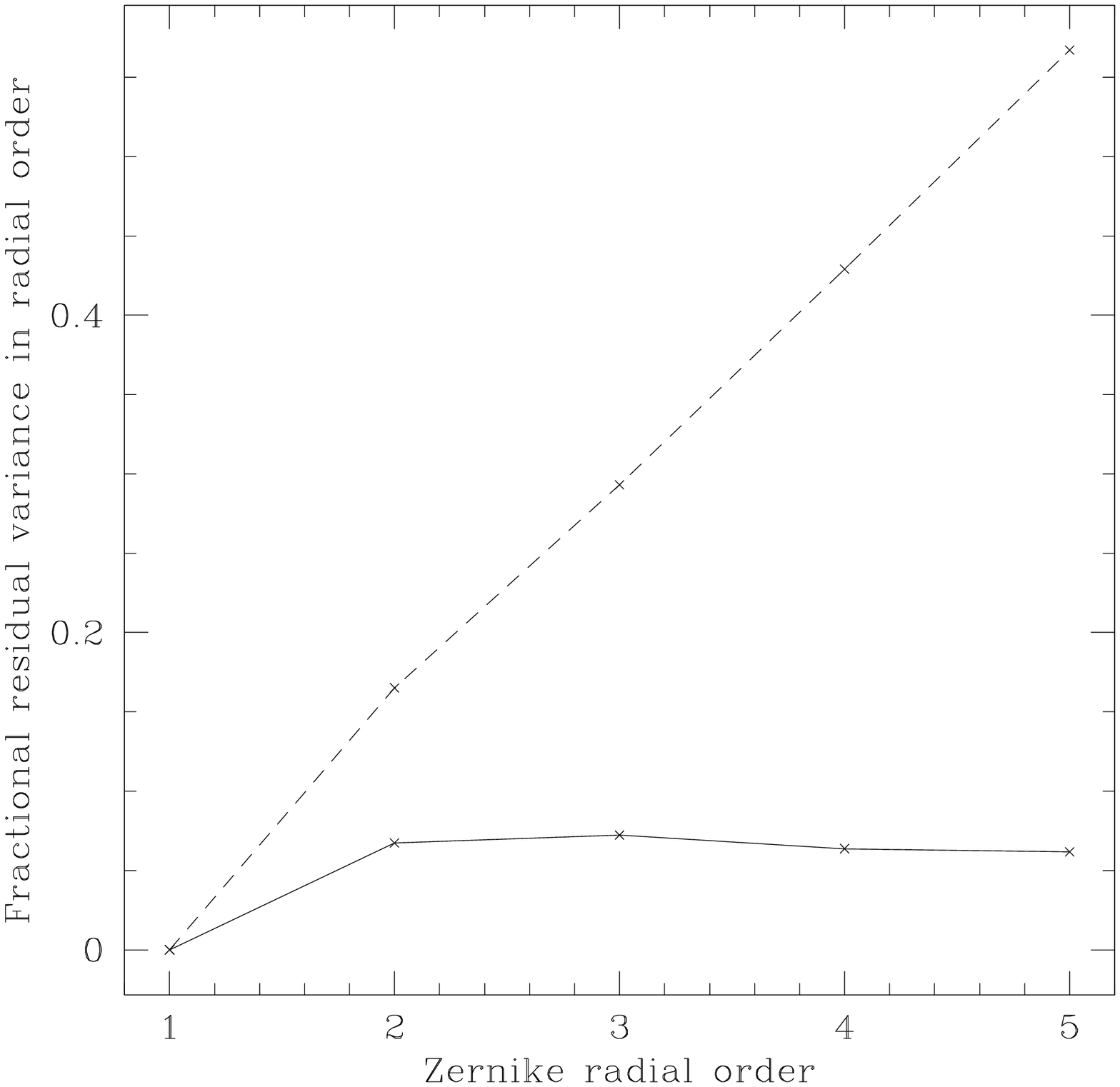}
\includegraphics[width=8cm]{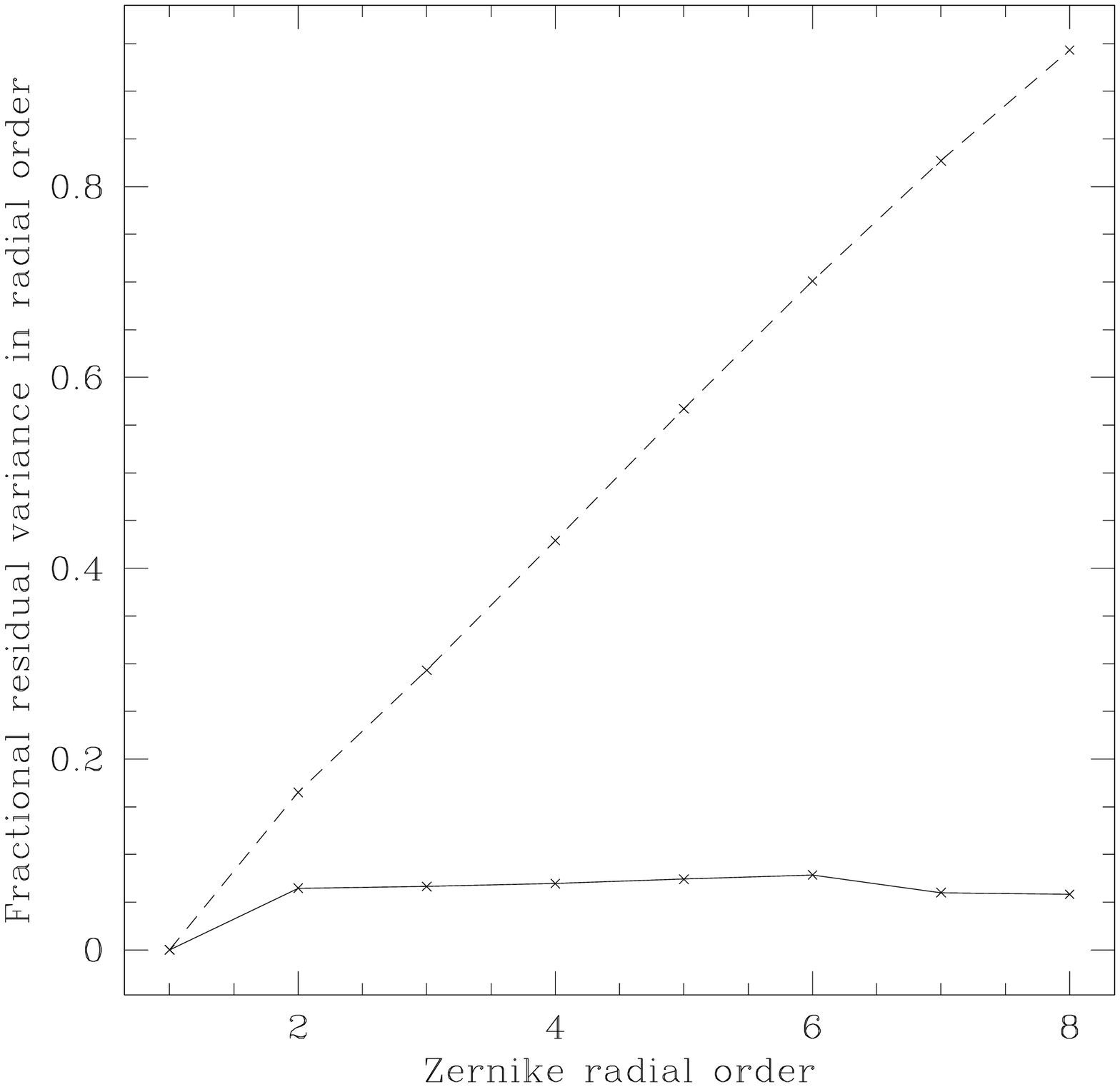}
\caption{Theoretical prediction of SPLASH performance (solid line) as compared with an equivalent conventional LGS/Shack-Hartmann wavefront sensor system (broken line). Results show residual wavefront variance as a fraction of uncorrected variance in each radial order of Zernikes, for a single atmospheric layer at 1/4 of the beacon altitude. \emph{Left:} $8 \times 8$ array of subapertures;  \emph{Right:} $12 \times 12$ array of subapertures.
}
\label{fig:theory_graph}
\end{figure*}

The matrices $\mathbf{G}$, $\mathbf{G}^{\prime}$ and $\mathbf{G}^{\prime \prime}$ are constructed by mapping tilt covariances for pairs of apertures with appropriate spatial separations into square arrays which use the same subaperture geometry along both axes as the subaperture axis of the $\curlyD$-matrices. An efficient method of calculating the tilt covariance between two spatially separated spatial apertures has been described by Ass\'emat \shortcite{assematthesis}. The tip/tilt covariances between subapertures $i$ and $l$ are given by
\begin{eqnarray}
	\langle g_{i,x} g_{l,x} \rangle & = & \frac{1}{2 d^2} \left[ \frac{\partial^2 D_{\varphi}}{\partial x^2} (x,y) \otimes I_{i l} (x,y) \right] \\
	\langle g_{i,x} g_{l,y} \rangle & = & \frac{1}{2 d^2} \left[ \frac{\partial^2 D_{\varphi}}{\partial x \partial y} (x,y) \otimes I_{i l} (x,y) \right]
\end{eqnarray}
where the $x$ and $y$ subscripts indicate the direction of the tilts. $I_{i l} (x,y)$ is the intercorrelation of the two aperture functions, defined by
\begin{equation}
	I_{i l} (x,y) = \int\!\!\!\int \Pi \left( \frac{u + x}{d} , \frac{v + y}{d} \right) \Pi \left( \frac{u}{d} , \frac{v}{d} \right) \ud u \ud v
\end{equation}
where $d$ is the subaperture size and $\Pi(x,y)$ is the aperture function, given by
\begin{equation}
	\begin{array}{cccl}
		\Pi(x,y) &= &1 & |x| < 1/2 \mbox{ and } |y| < 1/2 \\
		&= &0  & \mbox{ otherwise.}
	\end{array}
\end{equation}
$D_{\varphi}$ is the phase structure function which, assuming Kolmogorov turbulence with an infinitely large outer scale, is given by
\begin{equation}
	D_{\varphi} (x, y) = 6.88 r_0^{-5/3} (x^2 + y^2)^{5/6} .
\end{equation}
When one or both of the subapertures are affected by FA, one of the following equations is used:
\begin{equation}
	I_{i l}^{\prime} (x,y) = \int\!\!\!\int \Pi \left( \frac{u + x}{d} , \frac{v + y}{d} \right) \Pi \left( \frac{u}{\beta d} , \frac{v}{\beta d} \right) \ud u \ud v
\end{equation}
\begin{equation}
	I_{i l}^{\prime \prime} (x,y) = \int\!\!\!\int \Pi \left( \frac{u + \beta x}{\beta d} , \frac{v + \beta y}{\beta d} \right) \Pi \left( \frac{u}{\beta d} , \frac{v}{\beta d} \right) \ud u \ud v
\end{equation}
where $\beta$ is an FA factor, defined by
\begin{equation}
	\beta = 1 - \frac{h}{H} .
\end{equation}

Thus we have all the information required to predict the residual wavefront variance for a given number of corrected Zernike modes. We have assumed the cone geometry shown in Fig.~\ref{fig:splash}, and that the system is capable of perfectly correcting Zernike modes to the degree that they can be sensed. We also assume that tip and tilt across the full telescope aperture can be perfectly sensed, since in reality these modes would be sensed using a NGS rather than the laser beacon.

The results of the theoretical analysis of SPLASH are shown in Fig.~\ref{fig:theory_graph}. The benefits of the better sampling of the wavefront provided by SPLASH can clearly be seen in the plot. The fractional residual variance in each mode (i.e. the proportion of the variance in each mode that cannot be sensed and corrected) is roughly constant at each spatial scale for SPLASH, whereas the effect of FA on a conventional LGS becomes more severe as the spatial frequency of the aberrations increases. The results indicate that a SPLASH system could be expected to perform significantly better than an equivalent system using a conventional LGS with a Shack-Hartmann WFS, and that the benefits are greater for larger telescope apertures.

It is important to remember that we have ignored the effects of atmospheric aberrations on the downward path of the light from the focused spots. These effects are included in the numerical simulation described in the next section.

\section{CLOSED LOOP SIMULATION}
\label{sect:closed_sim}

\begin{figure*}%[htp]
\centering
\includegraphics[width=10cm]{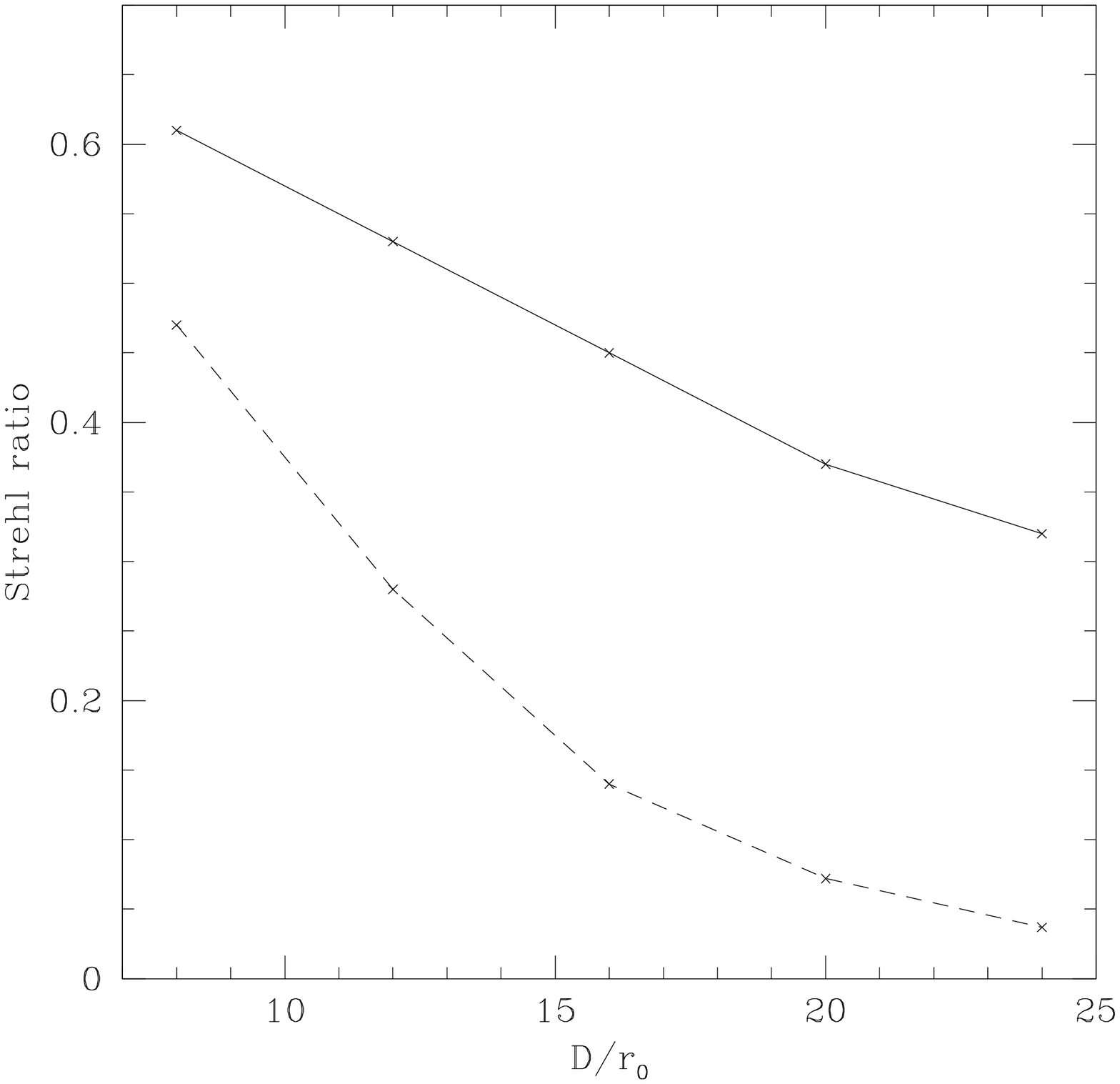}
\caption{Numerical simulation results: On-axis Strehl ratios for a SPLASH LGS AO system (solid line) and an equivalent conventional LGS AO system (broken line).}
\label{fig:strehl_graph}
\end{figure*}

A closed-loop semi-geometrical Monte Carlo simulation of SPLASH has also been implemented. This includes the effects of aberrations on the return path of the light through the atmosphere in addition to the upward-path FA considered in the theoretical analysis.

For the purposes of the simulation a uniform intensity distribution was assumed for the laser beams, with each beam being focused on the sky by a lens. The beacons were assumed to be sufficiently bright to ignore the effects of photon noise and CCD read noise.

The simulation assumes the diffraction-free geometry illustrated in figure \ref{fig:geometry} for the purposes of identifying the sections of atmosphere intersected by the beams, but far-field diffraction is included in the PSF calculations. For each sky spot, and at each time step, the phase aberration induced on the upward path is calculated and the on-sky PSF is calculated as a FFT of the complex amplitude across the subaperture. Similarly, the phase aberration on the downward path is projected onto the full telescope aperture and the downward-propagation PSF is calculated as the FFT of the complex amplitude across the telescope aperture. The total PSF (the LGS PSF imaged through the atmosphere) is calculated by convolving the upward and downward PSFs. 

The simulated spots are centroided separately - they are not ``stitched'' together into a complete spot pattern. It is assumed that if the combination of the size and motion of the spots is sufficient to cause cross-contamination, this can be compensated for by time-interleaving of the spots.

The wavefront corrector used is an idealised segmented deformable mirror (DM) consisting of square segments each capable of correcting piston, tip and tilt. Each mirror segment is aligned perfectly with a SPLASH subaperture. The DM segment tip and tilt values correspond directly to the corresponding spot $x$ and $y$ centroids, and the subaperture piston values across the DM are reconstructed from the centroids using a successive over-relaxation (SOR) algorithm \cite{southwell80}. As in the previous section, perfect (NGS) global tip/tilt correction is assumed.

The atmospheric model used consists of two translating Kolmogorov phase screens at different altitudes, with 60 per cent of the turbulence at the ground and the remaining 40 per cent at one quarter of the altitude of the laser beacons. A range of different $D/r_0$ values were simulated and for each one, the number of subapertures was matched to $D/r_0$, i.e. for $D/r_0 = 8$, an $8 \times 8$ subaperture WFS (and DM) was used. 

A conventional LGS AO system with the same DM, atmosphere model, beacon altitude, tip/tilt correction, etc. was also simulated to allow an objective comparison with SPLASH.

The simulations were carried out on a Cray XD1 supercomputer with 12 Opteron processors running SuSE linux. This machine features field programmable gate arrays (FPGAs) for hardware acceleration of numerically intensive processing, but the current version of the simulation code does not take advantage of these. Future versions are expected to do so \cite{basden}.

The control loop was successfully closed for all the $D/r_0$ values simulated and the SPLASH results compared favourably with the conventional LGS results, as can be seen in Figure~\ref{fig:strehl_graph}. SPLASH performs considerably better than a conventional LGS system across the whole range of simulated values. The highest $D/r_0$ values simulated correspond to an 8 m class telescope, so the technique could be suitable for use on existing telescopes.

As $D/r_0$ increases further, the residual wavefront error will worsen and centroid anisoplanatism become more noticeable.  For sufficiently large $D/r_0$ the residual wavefront error will prevent effective wavefront sensing, although the regime in which this occurs will be sensitive to the distribution of turbulence in the atmosphere. Further simulations to higher $D/r_0$ values (and with a range of atmosphere models) are required to assess the applicability of the SPLASH technique to ELTs.

\section{CONCLUSION}
\label{sect:conc}

We have described a new method of LGS wavefront sensing in which an array of Shack-Hartmann spots are projected onto the sky and then imaged through the telescope.

We have shown theoretically that, in the absence of any return-path wavefront aberrations, and assuming purely geometrical optics, such a system can be expected to suffer considerably less from FA than an equivalent conventional LGS system.

We have further demonstrated the validity of the technique using a semi-geometrical closed-loop simulation with a realistic atmosphere model, in which return-path aberrations were included in addition to upward-path turbulence. This simulation demonstrates the improvement in performance over a conventional single-LGS AO system for a range of $D/r_0$ values up to those approximately consistent with an 8 m class telescope.

We anticipate conducting further numerical simulation work for larger aperture sizes to investigate the feasibility of the SPLASH technique for ELTs. It may also be possible to experimentally verify the technique on-sky in the coming months.

\section{ACKNOWLEDGEMENTS}
\label{sect:ack}

Thanks to Fran\c cois Ass\'emat for drawing to our attention the method for calculating subaperture tilt covariances. We thank the referee, Celine d'Orgeville, for helpful comments which improved the paper. TB and RWW are grateful to the UK PPARC for financial support. This work forms part of a technology development programme (ELT Design Study) supported by the European Community within its Framework Programme 6, under contract No 011863.

\label{lastpage}

\end{document}